\documentclass[12pt,a4paper]{article}
\usepackage[utf8]{inputenc}
\usepackage[T1]{fontenc}
\usepackage{amsmath}
\usepackage{amssymb}
\usepackage{graphicx}
\usepackage{xcolor}
\usepackage{float}
\usepackage{caption}
\usepackage{subcaption}
\usepackage{array}
\usepackage{multirow}
\usepackage{longtable}
\usepackage{threeparttable}
\usepackage{booktabs}
\usepackage{tabularx}
\usepackage{geometry}
\usepackage{booktabs}
\usepackage{natbib}
\usepackage{hypernat}  \usepackage{hyperref}
\usepackage{soul}
\usepackage{svg}
\usepackage{enumitem}
\usepackage{colortbl}
\usepackage[capitalize,nameinlink]{cleveref}

\hypersetup{
    colorlinks=true,
    linkcolor=blue,
    filecolor=magenta,
    urlcolor=blue,
    citecolor=blue,       backref=true,         pdftitle={Skill-Driven Certification Pathways: Measuring Industry Training Impact on Graduate Employability},
    pdfauthor={Anatoli Kovalev, Narelle Stefanac, Marian-Andrei Rizoiu},
}

\title{Enhancing Educational Programs and Employment Outcomes through Industry-Based Certifications: A Framework for Measuring Impact}
\title{Bridging Educational Gaps Through Industry Certifications: A Data-Driven Framework for Quantifying Skill Alignment and Employment Readiness}
\title{From Classroom to Career: Detecting Skill Enhancement Patterns in Industry-Augmented Education Using Large-Scale Job Market Analysis}
\title{Skill-Driven Certification Pathways: Measuring Industry Training Impact on Graduate Employability}

\author{Anatoli Kovalev\\University of Technology Sydney \and Narelle Stefanac\\Microsoft \and Marian-Andrei Rizoiu\\University of Technology Sydney}
\date{}

\begin{document}

\maketitle

\begin{abstract}
Australia faces a critical technology skills shortage, requiring approximately 52,000 new technology professionals annually by 2030, while confronting a widening gap between employer requirements and graduate capabilities. With only 1\% of technology graduates considered immediately work-ready, traditional educational pathways alone prove insufficient to meet industry demands.
This research examines how industry certifications, such as Microsoft's AI-900 (Azure AI Fundamentals), can bridge this critical skills gap. We propose a novel, data-driven methodology that quantitatively measures skill alignment between educational offerings and job market requirements by analysing over 2.5 million job advertisements from Australia, the US, and the UK, mapping extracted skills to industry taxonomies using the Vectorised Skills Space Method.
Our findings reveal that combining university degrees with targeted industry certifications significantly enhances employability for technology roles. The Bachelor of Computer Science with AI major combined with AI-900 certification achieved the highest absolute skill similarity score for Machine Learning Engineer positions. 
Surprisingly, the largest improvements when augmented with AI certifications are experiences by non-technical degrees--such as nursing nursing--with up to $9,296\%$ percentage improvements in alignment with Machine Learning Engineer roles.
Our results challenge conventional assumptions about technology career pathways.
They can provide actionable insights for educational institutions seeking evidence-based curriculum design, students requiring strategic certification guidance, and employers recognising potential in candidates from non-traditional backgrounds who have obtained relevant certifications.

\end{abstract}

\section{Introduction}

Australia is facing a critical shortage of tech professionals, with The Australian Computer Society reporting that the nation is experiencing a skills shortage not seen in over six decades \citep{Sadler2023}. Based on current business projections, it is anticipated that by 2030, Australia will require a workforce of 1.3 million technology professionals, consequently, this demand corresponds to an annual need for approximately 52,000 technology workers \citep{ACS2024}. The shortage is exacerbated by rapid digital transformation and the increased adoption of emerging technologies across industries . In the 2021 report, titled ``Technology Council of Australia Roadmap to Deliver One Million Tech Jobs'', it has been suggested that the Australian technology sector stands as a cornerstone of the nation's economy, generating an annual contribution of \$167 billion to the Gross Domestic Product (GDP) and employing 861,000 individuals \citep{TCA2021}.

The traditional talent source has been evaluated by examining the number of graduates in technology fields from Australian universities and Vocational Education Training (VET) institutions. According to the ACS Digital Pulse report \citep{ACS2024}, only 75,000 ICT students entered the tech workforce in Australia between 2016 and 2021. That is equivalent to 15,000 ICT professionals joining the tech sector per year. Recently propose caps on the number of international students could potentially further jeopardise the future influx of technology workers. International students currently constitute 65\% of annual IT program completions \citep{ACS2024}.

Despite the high demand for tech workers, there is a significant mismatch between the skills required by employers and those possessed by the workforce. This mismatch results in many tech positions remaining unfilled, which in turn hampers business operations and growth. Recent research indicates that traditional learning pathways in Australia are facing challenges in producing job-ready workers. A 2019 study by Think Education revealed that nearly 40\% of young Australians feel their education has not provided them with the skills required for their current job \citep{Brown2019}. This sentiment is echoed in the Australian Workforce and Productivity Agency's report, which highlights industry perceptions that graduates are often ``not job ready'' \citep{Brown2019}.

Recent reports from key Australian technology organisations highlight significant challenges in producing job-ready graduates for the tech sector. The Australian Information Industry Association's (AIIA) State of the Nation 2024 report reveals a concerning trend in graduate work-readiness, with only 1\% of technology graduates considered immediately work-ready, down from 3\% in 2023 \citep{AssociationsForum2024}. The issue is further compounded by rising youth unemployment rates, reaching up to 20\% in some areas, and data showing that up to 30\% of university students are unemployed after completing their degrees \citep{Brown2019}.

The Australian Industry and Skills Committee (AISC) has recognised this gap, emphasising the need for a more integrated approach to education and workforce development. Their 2024 VET Workforce study projects that around 44\% of new jobs expected to be created over the next decade will have Vocational Education and Training (VET) as the primary training pathway \citep{JSA2024}. This underscores the importance of vocational skills training in addressing the national skills shortage and preparing workers for future job markets.

Furthermore, a comprehensive study on educational opportunity in Australia, published in 2020, reveals that approximately 29.7\% of 24-year-olds were not engaged in full-time education, training, or work \citep{Lamb2020}. This statistic suggests that the education system has not fully delivered on its promise to prepare all students in Australia for life, work, and learning. The study also highlights significant disparities based on socioeconomic status, gender, and geographic location, indicating that the challenges in producing job-ready workers are not uniformly distributed across the population \citep{Lamb2020}.

In response to these challenges, there is a growing recognition of the need for alternative pathways and more practical, work-oriented education models. The Australian Technical and Further Education (TAFE) system and other vocational education providers are increasingly seen as crucial in bridging the gap between traditional education and workforce needs \citep{JSA2024}. Additionally, initiatives like the National Skills Agreement (NSA) and fee-free TAFE programs are being implemented to increase access to vocational education and address specific skill shortages \citep{JSA2024}.

Universities are also increasingly integrating industry-specific training into their tech programs through various innovative approaches, aiming to produce graduates who are better prepared for the demands of the rapidly evolving tech sector. This includes implementing work-integrated learning (WIL) experiences, such as internships, placements, and industry-based projects \citep{BeanDawkins2021}. The National Priorities Industry Linkage Fund (NPILF), part of the Job-ready Graduates Package, aims to provide a framework for stepping up industry engagement in higher education \citep{BeanDawkins2021}. This initiative encourages universities to develop innovative WIL arrangements, including remote internships, placements, and industry-based projects. Furthermore, to provide more flexible and targeted learning options, universities are developing micro-credential offerings, often in collaboration with tech industry partners \citep{BeanDawkins2021}. These short courses allow students to acquire specific, in-demand skills quickly and provide a way for working professionals to upskill or reskill in emerging technologies \citep{BeanDawkins2021}. The Australian Government's introduction of the Undergraduate Certificate as a new qualification type demonstrates this shift towards shorter, more focused credentials \citep{BeanDawkins2021}.

The VET sector's approach to creating work-ready graduates offers valuable lessons for universities, particularly in the tech industry. By adopting and adapting strategies such as enhanced industry collaboration, flexible credentialing, digital skills integration, responsive program design, and a focus on employability skills, universities can better prepare their graduates for the dynamic tech sector. As the lines between vocational and higher education continue to blur, this cross-sector learning will be crucial in developing a workforce that can meet the evolving needs of the tech industry and drive innovation in the Australian economy.

\section{Background}

The dynamic nature of the technology sector has led to a growing disparity between the skills taught in traditional educational programs (Universities and TAFE) and those demanded by employers. This gap has given rise to the popularity of industry certifications as a means to bridge this divide. In today's rapidly evolving job market, the combination of traditional university degrees and industry-recognised certifications has become increasingly important for students seeking employment. This section of the paper examines the value of certifications from major technology companies such as Microsoft, AWS and Google when paired with a tertiary qualification. We will explore the impact of these combined credentials on employability, career advancement, and earning potential.

Over the past decade, there has been a significant increase in the popularity and recognition of industry-based certifications. These credentials, offered by technology companies like Microsoft, Amazon and Google, have become valuable assets for job seekers and professionals alike \citep{Certiport}. The growing demand for specialised skills in cloud computing, data analytics, and cybersecurity has further emphasised the importance of these certifications.

Research has shown that industry certifications can complement traditional university degrees by providing practical, job-ready skills that employers value. A study by the institute of Defence Analysis conducted in 2019 found that while employers generally preferred academic degrees, they also recognised the importance of certifications in identifying specific technical skills \citep{Belanich2019}. This suggests that the combination of a degree and relevant certifications can be particularly beneficial in the job market.

\subsection{Impact on Employability}

Several studies have demonstrated the positive impact of industry certifications on employability. According to a survey conducted by Amazon Web Services, 84\% of learners reported improved on-the-job efficiency after completing AWS training, while 83\% reported improved effectiveness \citep{AWS2022}. Additionally, 69\% of respondents reported higher earnings as a result of AWS training, and 74\% reported the same after earning AWS certifications \citep{AWS2022}.

A survey conducted by Coursera found that 88\% of learners believe an industry certificate will help them stand out with employers in the job market \citep{McAndrew}. This perception is validated by employer attitudes, as 92\% of employers surveyed said a certificate strengthens a candidate's application, and 76\% reported being more likely to hire a candidate with a certificate \citep{McAndrew}.

Microsoft Technical Certifications are powerful catalysts for employability, offering industry-recognised validation of job-ready skills. These certifications are tightly aligned with in-demand job roles and real-world scenarios, enabling individuals to demonstrate proficiency in critical technical domains. According to a 2024 IDC InfoBrief sponsored by Microsoft \citep{SmithMehta2024}, 70\% of organisations consider industry-recognised credentials essential or very important for skills training. Furthermore, a 2023 Pearson VUE Candidate Report~\citep{PearsonVUE} found that 91\% of certification holders felt more confident in their abilities, and 77\% reported being better able to innovate after earning their certification. Microsoft's credentials not only enhance professional credibility but also elevate visibility across networks, giving employers full confidence in the authenticity and relevance of certified skills. 

\subsection{Existing Methodologies for Researching Certification Impact}

Several methodologies have been employed to study the impact of industry certifications on employment outcomes:

\textbf{Correlation studies} examine the relationship between certification and various educational outcomes. For example, \citet{Cook2021} investigated the bivariate correlation between industry credentials earned during high school and post-graduation employment rates in related job areas. This methodology allows researchers to identify potential associations between certification and desired outcomes, such as employment rates or academic performance. However, it's important to note that correlation does not imply causation, and other factors may influence the observed relationships.
    
\textbf{Comparative Analysis:} \citet{Ramamonjiarivelo2020} used Welch t-tests to compare job satisfaction, and career growth between certified and non-certified individuals. This approach can be applied to compare outcomes for graduates with and without industry certifications.
    
\textbf{Surveys and Questionnaires:} Organisations like Global Knowledge conduct annual surveys to determine the average salaries associated with specific certifications \citep{GlobalKnowledge2024}. This methodology helps quantify the financial impact of certifications. While salary surveys provide concrete data on the monetary value of certifications, they may not capture other important aspects of career development and job satisfaction. A notable example of this approach is found in the study of Cisco Certified Network Associate (CCNA) courses. Researchers analysed data from the Waikato Institute of Technology (Wintec) and the Southern Institute of Technology (SIT) to evaluate how well the CCNA courses met the needs of the Information and Communications Technology (ICT) industry in Hamilton and Invercargill, New Zealand \citep{Rajendran2011}. The study compared the effectiveness of embedding CCNA in these regions and highlighted course topics found to be most useful in the workplace.

These methodologies offer diverse approaches to researching certification impact on educational programs. Researchers should carefully consider the strengths and limitations of each method when designing studies to evaluate the effectiveness of certification programs.

\section{A Novel Methodology for Assessing the Value of Industry Certifications in Conjunction with University Degrees}
\label{sec:methodology}

This section introduces our mathematical framework for measuring the impact of industry certifications on the skill alignment between university degrees and junior AI roles.
We introduce in \cref{subsec:skill-space-method} the prerequisites for our work, the Skill Space Method.
Next, we present a comprehensive methodology that quantifies skill relationships (\cref{subsec:skill-alignement}), combines skill sets from different educational components (\cref{subsec:combine-skillsets}), and evaluates the resulting improvement in job market alignment (\cref{subsec:quantify-certification-impact}).

\subsection{The Skills Space Method}
\label{subsec:skill-space-method}

To quantify skill alignment between degrees, certifications, and job roles, we use the Skills Space Method \citep{Dawson2021} as our foundation. 
This enables us to compute similarity scores between educational offerings and job requirements, specifically to evaluate certification impact.

The core of this method involves the following components:

\begin{enumerate}
    \item \textbf{Skill extraction:} Identifying skills from job advertisements, university curricula, and certification syllabi
    \item \textbf{Revealed Comparative Advantage (RCA):} Measuring the relative importance of skills within each educational or job context
    \item \textbf{Skill similarity matrix:} Quantifying relationships between different skills across educational and employment domains
    \item \textbf{Skill set similarity calculation:} Computing overall alignment between educational offerings and job requirements
\end{enumerate}

The Skill Set Similarity (SSS) between two skill sets A and B is calculated as:

\begin{equation} \label{eq:ssm}
SSS(A, B) = \frac{\sum_{i \in A, j \in B} w(i, A) \cdot w(j, B) \cdot \theta(i, j)}{\sqrt{\sum_{i \in A} w(i, A)^2} \cdot \sqrt{\sum_{j \in B} w(j, B)^2}}
\end{equation}

Where:
\begin{itemize}
    \item $w(i, A)$ is the weight of skill i in skill set A (e.g., a university degree)
    \item $w(j, B)$ is the weight of skill j in skill set B (e.g., a junior AI role)
    \item $\theta(i, j)$ is the similarity between skills i and j
\end{itemize}

\textbf{Temporal complexity and the Vectorised Skills Space Method.}
Note that the computation of the Skill Space Similarity in \cref{eq:ssm} has a complexity of $O(|A| \times |B|)$ for each pairwise comparison, where $|A|$ and $|B|$ are the sizes of the skill sets being compared. When applied to our large-scale analysis involving multiple degree programs, certifications, and job roles (see \cref{sec:analysis-data}), this results in computationally intensive operations that do not scale efficiently to real-world scenarios.
We contribute an efficient, vectorised implementation that we denote as the Vectorised Skills Space Method (VSSM) (fully detailed in \cref{app-subsec:vssm}). Through comprehensive benchmarking, VSSM achieves performance improvements of up to $143,876\times$ over naive implementations, enabling practical application to large datasets containing millions of job advertisements and thousands of unique skills.
This enables us to measure the baseline job alignment of university degrees and the enhanced alignment when industry certifications are added.

\subsection{Skill Alignment with Target Occupations}
\label{subsec:skill-alignement}

Beyond overall similarity, we need to identify which specific skills contribute most to employment readiness. 
We propose a granular skill alignment measurement function to determine how individual certification skills enhance job-specific competencies:

\begin{equation}
L(s_{target}, A, B) = \sum_{s \in A} w(s, A) \cdot w(s_{target}, B) \cdot \theta(s, s_{target}) \enspace,
\end{equation}
where:
\begin{itemize}
    \item $s_{target}$ is a specific job skill we want to measure alignment for
    \item A is the educational skill set (e.g., degree + certification)
    \item B is the job skill set (e.g., requirements for a Machine Learning Engineer)
    \item $w(s_{target}, B)$ is the importance of the target skill in the job
    \item $w(s, A)$ is the importance of a skill in the educational program
    \item $\theta(s, s_{target})$ is the similarity between an educational skill and the target job skill
\end{itemize}

This function identifies how well a combined education (university degree plus industry certification) prepares graduates for specific job requirements. By calculating this for key AI job skills, we can determine which certification-degree combinations best enhance employment prospects for specific roles.

\subsection{Combining Skill Sets from Degrees and Certifications}
\label{subsec:combine-skillsets}

To model how industry certifications complement university degrees, we formalise the integration of their respective skill sets.

Let us define:
\begin{itemize}
    \item $S_1$: University degree skill set (e.g., Bachelor of Computer Science)
    \item $S_2$: Industry certification skill set (e.g., Microsoft AI-900)
\end{itemize}

The combined skill set $S_1'$ is the union:

\begin{equation*}
    S_1' = S_1 \cup S_2 \enspace.
\end{equation*}

A crucial aspect is appropriately weighting skills in this combined set. 
Since certification skills are more recently acquired and often more aligned with current industry needs, we apply a linear transformation that positions these skills within the top $\alpha\%$ of the degree's weight range:
\begin{equation}
w^T(s, S_2) = a \cdot w(s, S_2) + b \enspace,
\end{equation}
with coefficients:
\begin{align*}
a &= \frac{\alpha \cdot (max_{S_1} - min_{S_1})}{max_{S_2} - min_{S_2}} \\
b &= (1-\alpha) \cdot (max_{S_1} - min_{S_1}) + min_{S_1} - a \cdot min_{S_2}
\end{align*}

Note that $\alpha$ is a hyperparamenter of our method, and in our analysis in \cref{sec:results-analysis} we set it to $\alpha=20\%$.
Where $max_{S_1}$, $min_{S_1}$, $max_{S_2}$, and $min_{S_2}$ are the maximum and minimum values of skill weights in the degree and certification skill sets, respectively:
\begin{align*}
min_{S_1} &= \min_{s \in S_1} w(s, S_1) \quad \text{(minimum weight in the degree skill set)} \\
max_{S_1} &= \max_{s \in S_1} w(s, S_1) \quad \text{(maximum weight in the degree skill set)} \\
min_{S_2} &= \min_{s \in S_2} w(s, S_2) \quad \text{(minimum weight in the certification skill set)} \\
max_{S_2} &= \max_{s \in S_2} w(s, S_2) \quad \text{(maximum weight in the certification skill set)}
\end{align*}

We compute the weights of each skill in the combined set as
\begin{equation}
w(s, S_1') = 
\begin{cases}
w(s, S_1) & \text{if } s \in S_1, s \notin S_2 \\
w^T(s, S_2) & \text{if } s \in S_2, s \notin S_1 \\
w(s, S_1) + w^T(s, S_2) & \text{if } s \in S_1 \cap S_2 \\
0 & \text{otherwise}
\end{cases}
\end{equation}

This formulation captures how industry certifications enhance university education: foundational degree skills maintain their importance, certification-specific skills receive weights reflecting their industry relevance, and overlapping skills gain enhanced importance—representing both theoretical foundations and practical applications.

\subsection{Quantifying Certification Impact on Job Alignment}
\label{subsec:quantify-certification-impact}

To measure precisely how industry certifications enhance employment readiness, we analyse the transformation in skill set similarity when certification skills are added to university degrees:

For a degree skill set $S_1$ and job role skill set $T$, the baseline similarity is:

\begin{equation}
    \Theta(S_1, T) = \frac{1}{C} \sum_{s_1 \in S_1} \sum_{t \in T} w(s_1, S_1) \cdot w(t, T) \cdot \theta(s_1, t)
\end{equation}

With normalisation constant $C = \sum_{s_1 \in S_1} \sum_{t \in T} w(s_1, S_1) \cdot w(t, T)$

For the certification-enhanced skill set $S_1'$:
\begin{equation}
    \Theta(S_1', T) = \frac{1}{C'} \sum_{s' \in S_1'} \sum_{t \in T} w(s', S_1') \cdot w(t, T) \cdot \theta(s', t)
\end{equation}

Through mathematical derivation (see the full derivation in \cref{app:math-derivations}), we can express this relationship as:
\begin{equation}
\Theta(S_1', T) = \frac{C}{C'} \cdot \Theta(S_1, T) + \frac{1}{C'} \sum_{s \in S_2} \sum_{t \in T} w^T(s, S_2) \cdot w(t, T) \cdot \theta(s, t)
\end{equation}

Where $C' = C + \sum_{s \in S_2} \sum_{t \in T} w^T(s, S_2) \cdot w(t, T)$

This equation reveals critical insights about certification impact:

\begin{itemize}
    \item When certification skills align well with job requirements ($\theta(s, t)$ is high), the combined similarity $\Theta(S_1', T)$ significantly exceeds the baseline degree-only similarity
    
    \item When certification skills have minimal relevance to target jobs ($\theta(s, t) \approx 0$), the combined similarity may actually decrease, as $\Theta(S_1', T) \approx \frac{C}{C'} \Theta(S_1, T) < \Theta(S_1, T)$ since $\frac{C}{C'} < 1$
\end{itemize}

This mathematical framework explains our empirical findings in \cref{sec:results-analysis} that non-technical degrees (like nursing) show dramatic improvement when combined with AI certifications, while technical degrees show more modest gains. It also explains why irrelevant certifications can theoretically reduce job alignment—the denominator ($C'$) increases without a corresponding increase in skill alignment.

This insight emphasises the importance of strategic certification selection aligned with specific career objectives and provides a theoretical basis for our measured improvements in degree-job alignment across different educational backgrounds.

\section{Analysis Framework and Data Sources}
\label{sec:analysis-data}

This section introduces the analysis framework and datasets required to quantify skill alignment between educational offerings and job market requirements.
\cref{subsec:analysis-framework} introduces the overall approach; next we introduce the data sources and their processing.
\cref{subsec:job-ads-dataset} introduces the online job ads dataset that captures the market requirements, and \cref{subsec:uni-degree-dataset,subsec:certifications-dataset} introduce, respectively, the industry certifications and university degrees datasets which capture the educational offers we consider.

\subsection{Analysis Framework}
\label{subsec:analysis-framework}

\begin{figure}[htbp]
    \centering
    \includegraphics[width=\textwidth]{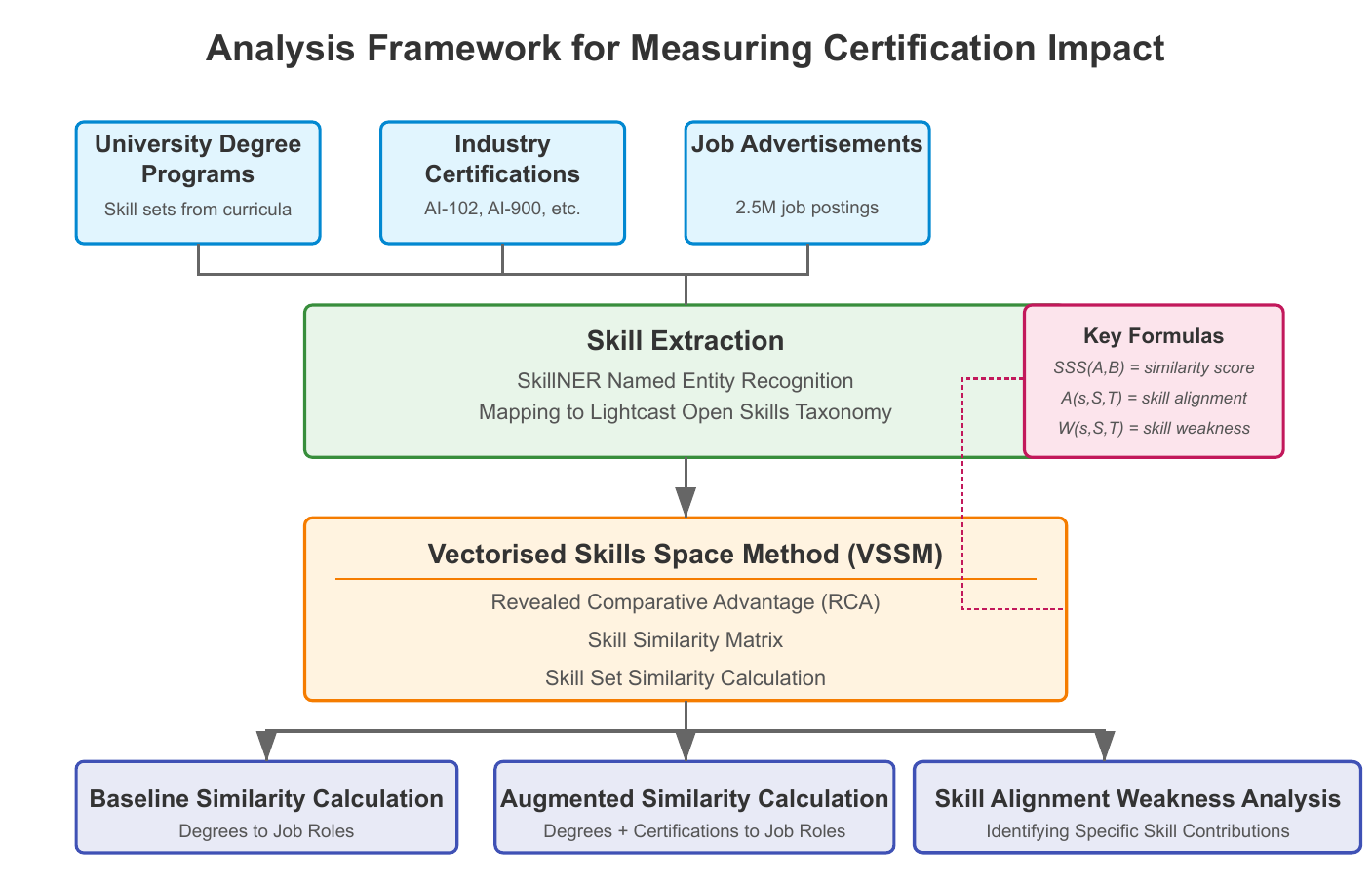}
    \caption{System architecture for skill alignment analysis and decision support.}
    \label{fig:system-architecture}
\end{figure}

\cref{fig:system-architecture} shows our methodology to quantify the impact of industry certifications when combined with university degrees. 
This approach allows for both theoretical assessment and real-world application.
The input data for the Skill Space Method (see \cref{subsec:skill-space-method}) comprises skills extracted from online job advertisements (market requirements) and educational offerings (industry certifications and university degrees). 
Our analysis follows a systematic process:

\begin{enumerate}
    \item \textbf{Baseline skill set establishment:} We first identify and quantify the skill sets provided by university degree programs, establishing a foundation for comparison.
    
    \item \textbf{Certification augmentation:} We then calculate the additional skills provided by industry certifications and modeled how these skills integrate with existing degree-based skills.
    
    \item \textbf{Job market alignment:} Using the Vector Skill Set Similarity (VSSM) calculation, we determine the alignment between these combined skill sets and contemporary job requirements.
    
    \item \textbf{Skill decomposition:} Finally, we identify specific skills contributing to alignment or misalignment, providing granular insights for curriculum development.
\end{enumerate}

\subsection{Data Source: Online Job Ads}
\label{subsec:job-ads-dataset}

We built and analysed a comprehensive dataset of $2,506,589$ job advertisements posted between June 2023 and June 2024. The dataset spans three economically developed English-speaking countries: Australia ($2,053,903$ listings), the United States ($410,982$), and the United Kingdom ($42,103$). 
This multi-national approach strengthens our analysis by capturing broader skill patterns across culturally and economically similar labour markets.

\paragraph{Skill Extraction Process.}
We employed a custom skill extraction pipeline built around the open-source SkillNER package to identify skills from job descriptions. 
SkillNER functions as a specialised rule-based natural language processing module designed specifically for extracting professional skills and certifications from unstructured text such as job postings and resumes \citep{AnasAito2022}. 
SkillNER leverages the SpaCy framework with large language models for initial named entity recognition, followed by semantic textual similarity techniques to map identified skill phrases to standardised terminology. 
Extracted skills are then aligned with the Lightcast Open Skills taxonomy, a comprehensive classification system containing over $33,000$ unique skills organised in hierarchical categories \citep{Lightcast2025}. 
This structured taxonomy, updated monthly to reflect emerging labour market trends, provides a standardised vocabulary for workforce analysis and enables precise comparison between educational offerings and industry requirements. 
From our dataset of 2.5 million job advertisements, we identified $11,210$ unique skills across various industries and experience levels. 
The job descriptions provided sufficient detail for effective skill extraction, with a median word count of $327$ (standard deviation = $289.65$, interquartile range = $142-522$).

\begin{figure}[tbp]
    \centering
    \includegraphics[width=\textwidth]{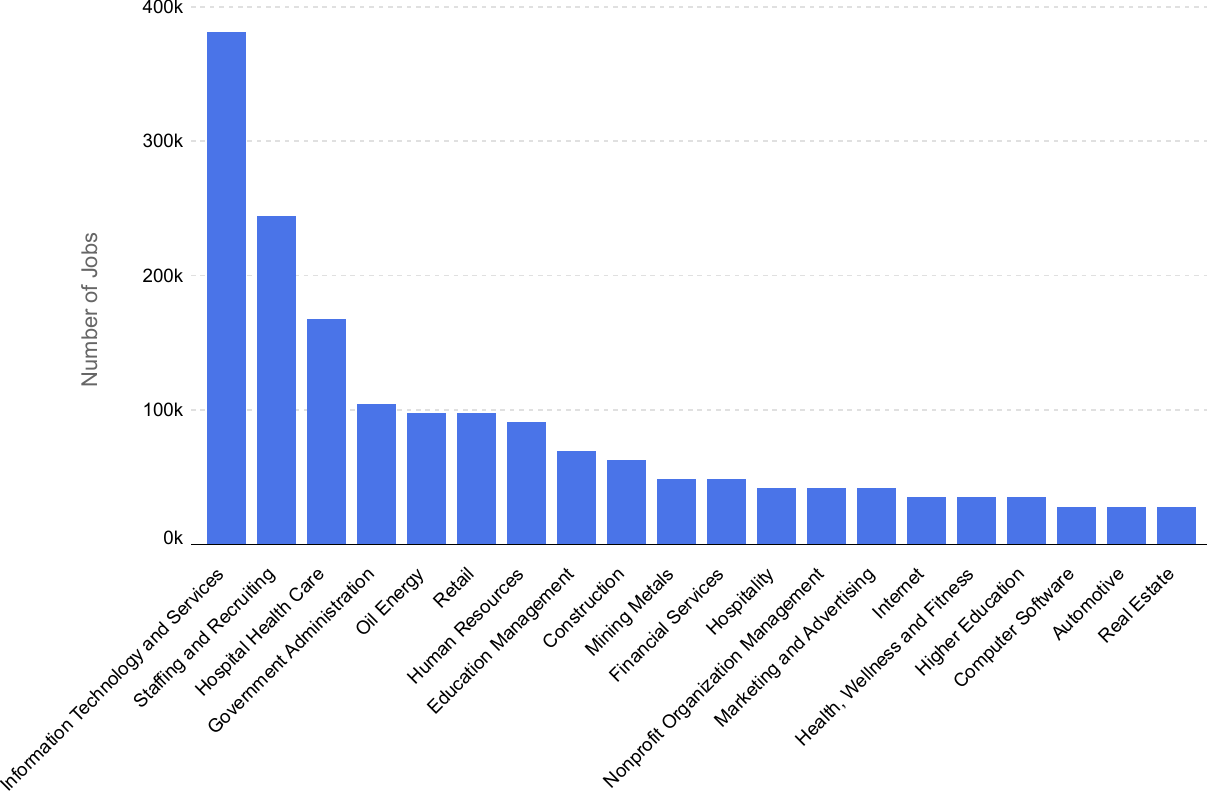}
    \caption{Top 20 Industry Sectors by Number of Job Advertisements}
    \label{fig:job-ads-per-sector}
\end{figure}

\paragraph{Dataset Profiling.}
The dataset exhibits diversity across industries and experience levels.
\cref{fig:job-ads-per-sector} illustrates the distribution across industry sectors, with Information Technology ($381,877$), Staffing and Recruiting ($178,133$), and Hospital \& Health Care ($122,607$) most prominently represented. The remaining sectors show a more uniform distribution ranging from approximately $19,000$ to $75,000$ advertisements each.

\begin{figure}[tbp]
    \centering
    \includegraphics[width=\textwidth]{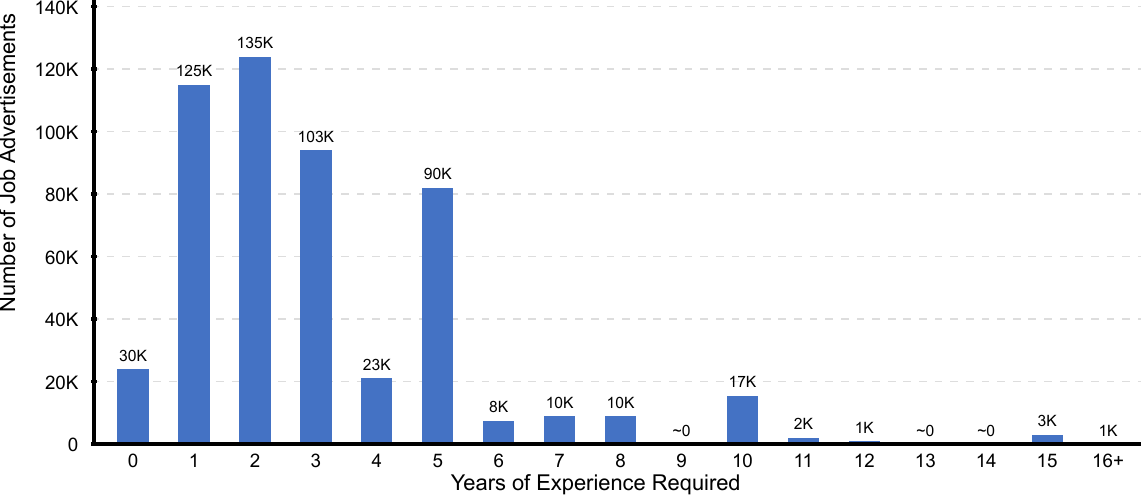}
    \caption{Distribution of Experience Requirements in Job Advertisements}
    \label{fig:job-ads-experience}
\end{figure}

Experience requirements, presented in \cref{fig:job-ads-experience}, demonstrate a balanced representation across career stages. 
Of the $564,584$ listings (22.52\% of total) that explicitly specified minimum experience requirements, most positions clustered around early to mid-career levels (1-5 years). 
The distribution is characterised by quartile values of $1.0$ (Q$_1$), $2.0$ (median), and $5.0$ (Q$_3$) years, with a mean of $3.08$ years slightly skewed by senior-level positions requiring 10+ years of experience.

For each job advertisement, we stored both extracted skills and relevant metadata, particularly job titles, enabling systematic grouping and comparison of skill sets across different roles. 
This structured approach facilitates the application of our skill similarity methodology to compare contemporary job market requirements against educational offerings detailed in the next sections.

\subsection{Data Source: University Degrees Programs}
\label{subsec:uni-degree-dataset}

We constructed a comprehensive dataset of degree programs and their constituent courses and subject from a medium-sized Australian university.
Each subject, course and degree is annotated with the skills that it provides to the students.

\subsubsection{Dataset Construction}

\paragraph{Educational offering structure.}
The educational offering follows a hierarchical model where degrees are composed of courses, and courses comprise individual subjects. 
Each subject includes lectures and practical tutorial sessions, with subjects potentially offered across multiple courses and degrees.
For example, the Bachelor of Computing Science (denoted as BCS) includes core subjects such as Discrete Mathematics, Programming 2, Data Structures and Algorithms, and Theory of Computing Science, alongside several specialised majors. 
The BCS Enterprise Software Development major (denoted as BCS (ESD)) encompasses subjects like Introduction to Software Development, Fundamentals of Interaction Design, Advanced Software Development, and Software Architecture, complemented by optional subjects such as Advanced Algorithms, Advanced Interaction Design, Advanced Internet Programming, Application Development in the iOS Environment, and Application Development with .NET.

\paragraph{Data collection and skill extraction.}
We systematically crawled the university's public-facing course handbook website, where each degree, course, and subject is described. 
For each academic offering, we extracted detailed information including: 
(1) subject descriptions and learning outcomes, 
(2) availability schedules (semester offerings), 
(3) learning and teaching activities, 
(4) assessment methodologies, 
(5) prerequisites and corequisites, and 
(6) additional subject-specific details.

We use SkillNER (see \cref{subsec:job-ads-dataset}) to extract skills from the description of each subject, interpreting these as the competencies taught within that subject. 
The complete skill set for each degree program was constructed as the union of skills present in: 
(1) core subjects required for degree completion, 
(2) core subjects within the chosen major specialisation, and 
(3) all available elective subjects.

We note that while students typically select electives up to a specified credit point limit (24 credit points for BCS (ESD)), we included all elective options in our analysis as we lack access to enrollment statistics to predict student choices. 
Our investigation revealed that elective subjects typically demonstrate high skill overlap, mitigating potential bias from this methodological choice.

\textbf{Degree programs.}
For this work, we selected four representative undergraduate degree programs:
\begin{enumerate}[leftmargin=3cm, labelwidth=6em, align=left]
    \item[\textbf{BCS (AI)}] Bachelor of Computing Science with Artificial Intelligence and Data Analytics major
    \item[\textbf{BCS (ESD)}] Bachelor of Computing Science with Enterprise Software Development major
    \item[\textbf{BEng}] Bachelor of Engineering (Honours) with Mechanical Engineering major
    \item[\textbf{BNur}] Bachelor of Nursing
\end{enumerate}

We chose these programs to examine certification impact across a diverse educational spectrum: two programs within information technology, one adjacent STEM discipline (engineering), and one completely distinct field (nursing).

\subsubsection{Dataset Profiling}

\begin{figure}[tbp]
    \centering 
    \newcommand\myheight{0.23}
    \subfloat[]{
        \includegraphics[height=\myheight\textheight]{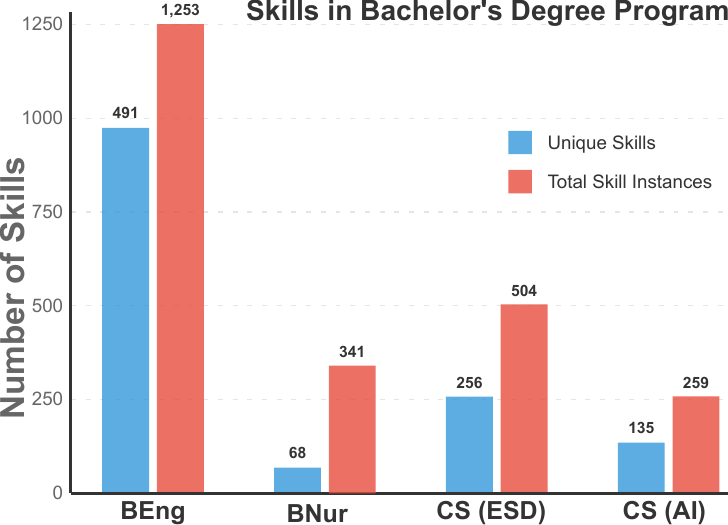}\label{fig:skills-comparison}
    }
    \subfloat[]{
        \includegraphics[height=\myheight\textheight]{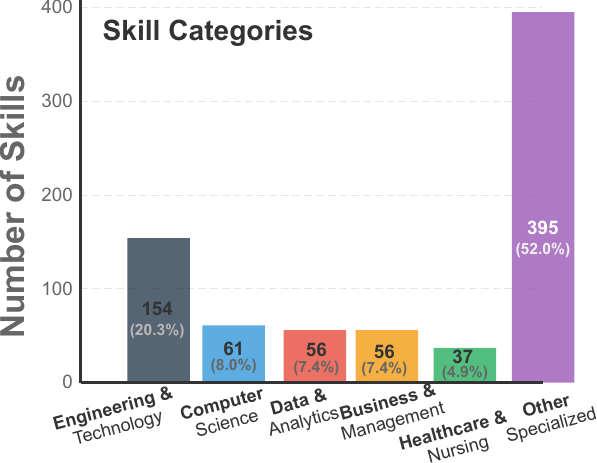}
        \label{fig:skill-categories}
    }
    \caption{
        \textbf{Skill analysis across degree programs and categories. }
        (a) Comparison of skill counts across different bachelor's degree programs demonstrates the substantial variation in skill set sizes, with Engineering having the largest and Nursing the smallest skill sets. 
        (b) Distribution of skills across major categories reveals that over half of all skills (52.0\%) fall into specialised domains beyond the main technical categories.
    }
    \label{fig:skill-analysis}
\end{figure}

Our dataset encompasses significant diversity in both scope and specialisation across the four degree programs, as illustrated in \cref{fig:skill-analysis}.

\paragraph{Dataset Overview.}
As shown in \cref{fig:skills-comparison}, the degree programs demonstrate substantial variation in skill set sizes. 
The Bachelor of Engineering contains the largest skill repository with 491 unique skills across $1,253$ total instances, while the Bachelor of Nursing represents the most focused program with $68$ unique skills distributed across $341$ instances. 
The two Computer Science programs occupy an intermediate position, with Enterprise Software Development encompassing $256$ unique skills ($504$ total instances) and the Artificial Intelligence specialisation containing $135$ unique skills ($259$ total instances).

Aggregate statistics reveal
(1) $759$ total unique skills across all programs, 
(2) $2,357$ total skill instances (including repetitions), 
(3) $589.3$ average skills per program, and (4) $145$ skills ($19.1\%$) appear in multiple programs (cross-program skill overlap).

\paragraph{Skill Repetition Patterns.}
The dataset exhibits varying levels of skill duplication within programs, ranging from $47.9\%$ (CS AI) to $80.1\%$ (Nursing), reflecting curricular emphasis and structural differences across disciplines. 
The substantial variation in skill set sizes shown in \cref{fig:skills-comparison} demonstrates how different programs emphasise skill breadth versus depth. Higher repetition rates in nursing indicate concentrated skill reinforcement in core competencies, while lower rates in computer science suggest broader skill diversity across the curriculum.

\paragraph{Domain Distribution.}
\cref{fig:skill-categories} presents the categorical distribution of skills across major domains. 
The analysis reveals that over half of all skills ($52.0\%$, $395$ skills) fall into specialised domains beyond the main technical categories, highlighting the diverse nature of modern degree programs. 
The primary technical categories show: 
Engineering \& Technology leads with $154$ skills ($20.3\%$), followed by 
Computer Science with $61$ skills ($8.0\%$), 
Data \& Analytics with $56$ skills ($7.4\%$), 
Business \& Management with $56$ skills ($7.4\%$), and 
Healthcare \& Nursing with $37$ skills ($4.9\%$). 
This distribution underscores the multidisciplinary nature of contemporary higher education, where technical skills are complemented by substantial components from business, communication, and domain-specific areas.

\paragraph{Technology Specificity}
The dataset includes only 10 technology-specific skills (1.3\%), referencing particular platforms, tools, or vendors (Microsoft, IBM, iOS, HTML, etc.), providing granular detail for technical competencies essential in contemporary professional environments. 
This low proportion of vendor-specific skills indicates that university programs maintain a platform-agnostic approach, concentrating on fundamental technologies and foundational approaches rather than training students on specific commercial tools or proprietary systems.

\paragraph{Program-Specific Skill Profiles.}

Each degree program demonstrates distinct skill emphases aligned with their respective domains:

\begin{itemize}
    \item \textbf{BEng:} Predominantly technical skills ($25.5\%$ engineering-specific, $8.8\%$ business \& management)
    \item \textbf{BNur:} Highly specialised healthcare focus ($50.0\%$ healthcare-specific skills)
    \item \textbf{BCS (ESD):} Balanced technical profile ($17.2\%$ engineering/technology, $13.3\%$ computer science)
    \item \textbf{BCS (AI):} Data-centric specialisation ($17.0\%$ data/analytics, $11.9\%$ engineering \& technology)
\end{itemize}

\paragraph{Cross-Program Skill Overlaps.}
Our analysis reveals $145$ skills shared across multiple programs, with the most frequent being foundational competencies, such as
Problem Solving ($94$ instances across programs),
Decision Making ($85$ instances),
FourGen Computer-Aided Software Engineering (CASE) Tools ($70$ instances),
Learning Strategies ($64$ instances), and 
System Requirements ($48$ instances).

The Jaccard similarity analysis confirms minimal skill overlap between programs, supporting our hypothesis that degrees provide largely distinct skill sets.
BCS (AI) and BCS (ESD) have only $15.0\%$ skill overlap. 
This indicates that even within the same discipline, different tracks provide largely distinct competencies. 
While some foundational skills (Information Technology, Computer Science, Problem Solving, and Artificial Intelligence) are shared, the majority of skills are specialised to each track's focus area.
Other programs pairs show similar trends: BEng and BCS (ESD) have $17.8\%$ overlap, while BEng and BCS (AI) have $9.2\%$ overlap.
Finally, BNur is distinct from all other programs with $<3.2\%$ similarity

\subsection{Data Source: Microsoft Industry Certifications}
\label{subsec:certifications-dataset}

For the purpose of this work, we focus on the Microsoft certification AI-900 (Azure AI Fundamentals). The Microsoft Azure AI-900 certification is designed to validate an individual's skills and knowledge in artificial intelligence (AI) and machine learning (ML) technologies on the Microsoft Azure platform. This certification caters to foundation level of expertise and focus on distinct aspects of AI implementation.

\begin{table}[h]
    \centering
    \begin{tabular}{ll}
        \toprule
        \textbf{Aspect}                       & \textbf{AI-900}                \\
        \midrule
        \textbf{Level}                        & Beginner                   \\
        \textbf{Target Audience}              & Technical and non-technical roles \\
        \textbf{Focus}                        & Broad overview of AI concepts  \\
        \textbf{Depth of Technical Knowledge} & Basic                          \\
        \textbf{Hands-on Experience Required} & Minimal                        \\
        \bottomrule
    \end{tabular}
    \caption{Microsoft Azure AI-900 Certification Overview}
\end{table}

The Microsoft Azure AI Fundamentals certification (AI-900) is a foundational-level course designed to introduce learners to artificial intelligence (AI) concepts and the various AI services available through Microsoft Azure \citep{K21Academy2025}. This certification is particularly valuable for individuals seeking to enhance their understanding of AI without requiring extensive technical expertise, making it accessible to a broad audience including students, business professionals, IT experts, and developers \citep{CertEmpire2025}.

\paragraph{Accessibility and Prerequisites}
One significant advantage of the AI-900 certification is its accessibility; it has no formal prerequisites. However, basic familiarity with cloud computing concepts can be beneficial for participants \citep{CertEmpire2025}. This makes the course ideal for non-technical professionals who wish to develop foundational knowledge in AI technologies.

\paragraph{Benefits of the AI-900 Certification}

The Microsoft Azure AI Fundamentals certification offers several key benefits:

\begin{enumerate}
    \item \textbf{Foundational Knowledge:} It provides participants with a solid grounding in essential AI concepts which can serve as a stepping stone towards more advanced certifications such as Azure AI Engineer Associate (AI-102) or Azure Data Science Associate (DP-100)
    
    \item \textbf{Industry Recognition:} Microsoft certifications are widely recognised by employers globally, enhancing employability by demonstrating verified knowledge in AI fundamentals
    
    \item \textbf{Career Opportunities:} Even at an entry-level certification stage, it opens opportunities in roles such as AI business analysts or cloud consultants where an understanding of AI principles is beneficial
    
    \item \textbf{Accessibility for Non-Technical Professionals:} The absence of prerequisites makes it suitable for professionals from diverse backgrounds who aim to incorporate AI into their business operations or career paths
    
    \item \textbf{Hands-on Experience with Azure Services:} Participants gain practical exposure to Azure's leading-edge technologies including Machine Learning services, Cognitive Services for NLP and computer vision tasks, and conversational agents like chatbots
\end{enumerate}

\paragraph{Relevance to Educational Programs and Employment Outcomes}

In the context of enhancing educational programs and employment outcomes through industry-based certifications---a primary focus of our research---the Microsoft Azure AI Fundamentals (AI-900) certification aligns effectively by providing measurable impacts on employability and foundational skill development. According to recent insights from Microsoft Education initiatives, incorporating certifications like the AI-900 significantly contributes towards preparing an AI-skilled workforce capable of thriving in an increasingly technology-driven job market \citep{Microsoft2025}. Such credentials not only validate proficiency but also enhance employment opportunities by equipping learners with industry-recognised skills that employers value.

\section{Results and Analysis}
\label{sec:results-analysis}

This section presents our empirical findings on the impact of industry certifications when combined with university degrees.
It applies the methodology established in \cref{sec:methodology} to the datasets described in \cref{sec:analysis-data}. 
We begin by establishing baseline skill alignment measurements between university degrees and junior AI roles (\cref{subsec:baseline-similarity}), providing the foundation for subsequent impact analysis. We then quantify how industry certifications enhance these alignments (\cref{subsec:cert-impact}), revealing both expected improvements and surprising patterns across different educational backgrounds. 
Finally, we investigate an unexpected phenomenon whereby non-technical degrees experience dramatic improvements when augmented with AI certifications (\cref{subsec:nursing-anomaly}), challenging conventional assumptions about technology career pathways and skill development strategies.

\subsection{Baseline Skill Set Similarity}
\label{subsec:baseline-similarity}

This section establishes baseline measurements by computing the Skill Set Similarity (\cref{eq:ssm}) between our four selected university degrees and representative junior AI roles. To provide context, we also include the skill similarity of the AI-900 certification for direct comparison. We select six representative roles spanning AI and software engineering that require minimal prior expertise, as illustrated in \cref{fig:job-ads-experience}. We use our Vectorised Skills Space Method (VSSM) computation (see \cref{subsec:skill-space-method}), and we present the results in \cref{tab:baseline-similarity}.

\begin{table}[htbp]
    \centering
    \caption{
        The Skill Set Similarity (see \cref{eq:ssm}) of Undergraduate Degrees and AI-900 (in columns) to Junior AI Roles (in lines).
        The background color indicate the range between lowest similarity (white background) to highest similarity (blue).
        All values should be read multiplied by $10^{-4}$.
    }
    \label{tab:baseline-similarity}
    \begin{tabular}{lcccc|c}
        \toprule
        \textbf{Role}             & \textbf{BNur}          & \textbf{BEng}          & \textbf{BCS (ESD)}       & \textbf{BCS (AI)}        & \textbf{AI-900}           \\
        \midrule
        Software Engineer         & \cellcolor{blue!2}1.96 & \cellcolor{blue!9}8.50 & \cellcolor{blue!25}21.70 & \cellcolor{blue!16}13.71 & \cellcolor{blue!21}18.61  \\
        AI Software Engineer      & \cellcolor{blue!0}0.80 & \cellcolor{blue!5}4.74 & \cellcolor{blue!11}9.69  & \cellcolor{blue!8}6.81   & \cellcolor{blue!13}11.21  \\
        Data Engineer             & \cellcolor{blue!3}2.60 & \cellcolor{blue!6}5.72 & \cellcolor{blue!15}13.07 & \cellcolor{blue!17}15.24 & \cellcolor{blue!43}36.73  \\
        Data Scientist            & \cellcolor{blue!0}0.39 & \cellcolor{blue!4}4.18 & \cellcolor{blue!11}9.63  & \cellcolor{blue!12}10.91 & \cellcolor{blue!46}39.96  \\
        Machine Learning Engineer & \cellcolor{blue!0}0.53 & \cellcolor{blue!2}2.21 & \cellcolor{blue!6}5.88   & \cellcolor{blue!56}48.13 & \cellcolor{blue!100}85.87 \\
        Data Analyst              & \cellcolor{blue!2}2.27 & \cellcolor{blue!4}4.14 & \cellcolor{blue!7}6.59   & \cellcolor{blue!12}10.39 & \cellcolor{blue!10}9.03   \\
        \bottomrule
    \end{tabular}
\end{table}

Our analysis of undergraduate degrees reveals varied alignment with junior AI roles.
The Bachelor of Computer Science with AI major (BCS (AI)) produces the highest skill set similarity to any junior AI role, achieving a similarity score of $48.13$ with Machine Learning Engineer positions. This result aligns with expectations, as the AI major specifically targets data and machine learning competencies that directly correspond to contemporary industry requirements.

The data reveal distinct specialisation patterns across degree programs. For data-focused and machine learning roles (Data Engineer, Data Scientist, Data Analyst, and Machine Learning Engineer), BCS (AI) consistently demonstrates the strongest alignment. 
Conversely, for more generalised software development positions (Software Engineer and AI Software Engineer), BCS (ESD) exhibits superior skill similarity. This differentiation reflects the targeted nature of university specialisations and their alignment with specific career pathways.

The Bachelor of Engineering with Mechanical major (BEng) occupies an intermediate position across all roles, demonstrating moderate skill similarity that reflects the mathematical and analytical foundations common to engineering disciplines. As anticipated, the Bachelor of Nursing shows the lowest overall skill set similarity across all AI roles, which is intuitive given its primary focus on biological sciences and medical practice rather than the computational and statistical foundations prevalent in technology roles.

A surprising finding emerges when examining the AI-900 certification in isolation. This industry certification alone demonstrates higher skill similarity to several junior AI roles than many complete undergraduate degrees. For Machine Learning Engineer positions, AI-900 achieves a similarity score of $85.87$, exceeding even BCS (AI)'s score of $48.13$. Similarly, for Data Scientist roles, AI-900 outperforms all undergraduate degrees with a score of $39.96$, compared to BCS (AI)'s score of $10.91$. For Data Engineer positions, AI-900 records a similarity of $36.73$, surpassing all degrees except the BCS with AI major.

This finding proves particularly significant as it demonstrates that a focused industry certification can potentially provide more relevant skills for specific AI roles than entire undergraduate degree programmes. The result underscores the value of targeted certification pathways, especially for professionals seeking to transition into specialised AI careers. However, it is important to acknowledge that direct comparison between a targeted industry certification and an entire undergraduate degree programme is not entirely equitable. Bachelor's degrees, particularly in computing disciplines, must devote substantial curriculum to fundamental skills such as mathematics, algorithms, and theoretical computer science. While these foundational skills may not appear directly in job requirements for junior roles, they constitute necessary dependencies for more specialised skills and provide the conceptual groundwork for long-term career development. Industry certifications, by design, can focus exclusively on immediately applicable skills precisely because they build upon this foundational knowledge, rather than needing to establish it from first principles.

\subsection{Impact of Industry Certifications}
\label{subsec:cert-impact}

This section measures the impact of adding industry certifications to university degrees on skill alignment with junior AI roles. We apply our proposed approach for combining skill sets (see \cref{subsec:combine-skillsets}) to quantify the effect of augmenting each of the four university degrees with the AI-900 certification. The enhanced skill similarities are presented in \cref{tab:cert-impact}, with percentage improvements detailed in \cref{tab:cert-percentage}.

\begin{table}[htbp]
    \centering
    \caption{Skill set similarity between roles (in lines) and the Undergraduate Degrees enhanced with the AI-900 certification.}
    \label{tab:cert-impact}
    \begin{tabularx}{\textwidth}{lXXXX}
        \toprule
        \textbf{Role}             & \textbf{BNur + AI-900} & \textbf{BEng + AI-900} & \textbf{BCS (ESD) + AI-900} & \textbf{BCS (AI) + AI-900} \\
        \midrule
        Software Engineer & \cellcolor{green!9}10.66 & \cellcolor{green!13}13.06 & \cellcolor{green!33}23.72 & \cellcolor{green!27}20.17 \\
        AI Software Engineer & \cellcolor{green!0}6.27 & \cellcolor{green!2}7.35 & \cellcolor{green!10}11.40 & \cellcolor{green!9}11.07 \\
        Data Engineer & \cellcolor{green!28}20.84 & \cellcolor{green!20}16.52 & \cellcolor{green!32}23.16 & \cellcolor{green!41}27.98 \\
        Data Scientist & \cellcolor{green!26}19.78 & \cellcolor{green!13}13.21 & \cellcolor{green!20}16.80 & \cellcolor{green!28}21.10 \\
        Machine Learning Eng. & \cellcolor{green!83}49.80 & \cellcolor{green!48}31.36 & \cellcolor{green!65}40.51 & \cellcolor{green!100}59.01 \\
        Data Analyst & \cellcolor{green!0}6.09 & \cellcolor{green!0}6.34 & \cellcolor{green!5}8.87 & \cellcolor{green!10}11.37 \\
        \bottomrule
    \end{tabularx}
\end{table}

\begin{table}[htbp]
    \centering
    \caption{Percentage Increase in Skill Set Similarity with AI-900}
    \label{tab:cert-percentage}
    \begin{tabular}{lcccc}
        \toprule
        \textbf{Role}             & \textbf{BNur}             & \textbf{BEng}            & \textbf{BCS (ESD)}      & \textbf{BCS (AI)}      \\
        \midrule
        Software Engineer         & \cellcolor{red!56}444\%   & \cellcolor{red!26}54\%   & \cellcolor{red!0}9\%    & \cellcolor{red!24}47\% \\
        AI Software Engineer      & \cellcolor{red!62}684\%   & \cellcolor{red!26}55\%   & \cellcolor{red!10}18\%  & \cellcolor{red!28}63\% \\
        Data Engineer             & \cellcolor{red!63}702\%   & \cellcolor{red!44}189\%  & \cellcolor{red!31}77\%  & \cellcolor{red!32}84\% \\
        Data Scientist            & \cellcolor{red!91}4972\%  & \cellcolor{red!46}216\%  & \cellcolor{red!30}74\%  & \cellcolor{red!34}93\% \\
        Machine Learning Engineer & \cellcolor{red!100}9296\% & \cellcolor{red!72}1337\% & \cellcolor{red!60}589\% & \cellcolor{red!14}23\% \\
        Data Analyst              & \cellcolor{red!42}168\%   & \cellcolor{red!26}53\%   & \cellcolor{red!20}35\%  & \cellcolor{red!0}9\%   \\
        \bottomrule
    \end{tabular}
\end{table}

\cref{tab:cert-impact} quantifies how Microsoft's AI-900 certification affects skill matching when added to undergraduate degrees. As expected, BCS (AI) plus AI-900 achieved the highest absolute score at $59.01 \times 10^{-4}$ for Machine Learning Engineer roles, which is logical given that adding AI-specific certifications to AI-focused undergraduate training should yield the strongest alignment with AI-specific positions.

However, \cref{tab:cert-percentage} reveals a more nuanced and intriguing story by examining the percentage increase in skill set similarity after adding the AI-900 certification to undergraduate degrees (compared to the baseline similarities shown in \cref{tab:baseline-similarity}). Nursing graduates experienced extraordinary improvements: $9296\%$ for Machine Learning Engineer, $4972\%$ for Data Scientist, and $702\%$ for Data Engineer roles. We analyse this ``nursing anomaly'' in detail in \cref{subsec:nursing-anomaly}. In contrast, technical degrees showed more modest gains---BCS (AI) improved by only $23\%$ for Machine Learning Engineer roles despite achieving the highest absolute score.

A clear pattern emerges across role types. AI-focused positions (Machine Learning Engineer, Data Scientist) benefited substantially more from AI-900 certification than traditional software development roles. This observation, combined with the alignment scores in \cref{tab:baseline-similarity}, suggests the certification is particularly well-suited to these emerging technical positions. Software engineering roles showed smaller improvements because computer science degrees already provide substantial relevant skills, and because AI-900 is less tailored towards these established roles.

\begin{table}[tbp]
    \centering
    \caption{Top Skills by Alignment Score for Bachelor of Nursing + AI-900 by Junior AI Role}
    \label{tab:nursing-skills}
    \begin{tabular}{lcc}
        \toprule
        \textbf{Role} & \textbf{Skill} & \textbf{Alignment Score} \\
        \midrule
        \multirow{5}{*}{Software Engineer} & Application Programming Interface (API) & 25.77 \\
         & C\# Programming Language & 25.76 \\
         & Software Development & 15.43 \\
         & Back End Software Engineering & 5.69 \\
         & Systems Integration & 4.31 \\
        \midrule
        \multirow{5}{*}{Data Engineer} & Azure Machine Learning & 28.9 \\
         & Data Store & 7.23 \\
         & Data Discovery & 7.21 \\
         & Knowledge Graph & 6.92 \\
         & Source Data & 6.49 \\
        \midrule
        \multirow{5}{*}{ML Engineer} & Azure Machine Learning & 222.91 \\
         & Gradient Boosting & 107.23 \\
         & Natural Language Processing & 73.58 \\
         & Text Mining & 46.2 \\
         & Topic Modeling & 41.16 \\
        \midrule
        \multirow{5}{*}{AI Software Engineer} & C\# Programming Language & 16.94 \\
         & Information Retrieval & 7.69 \\
         & Software Development & 6.24 \\
         & Artificial Intelligence & 4.05 \\
         & Code Review & 3.59 \\
        \midrule
        \multirow{5}{*}{Data Analyst} & Applied Mathematics & 7.53 \\
         & Extract Transform Load (ETL) & 5.3 \\
         & Power BI & 1.72 \\
         & Cardiac Monitoring & 1.55 \\
         & Data Analysis & 1.46 \\
        \midrule
        \multirow{5}{*}{Data Scientist} & Machine Learning & 19.55 \\
         & Linear Regression & 15.24 \\
         & Data Science & 11.09 \\
         & Artificial Intelligence & 8.94 \\
         & Algorithmic Trading & 6.62 \\
        \bottomrule
    \end{tabular}
\end{table}

\subsection{The "Nursing Anomaly"}
\label{subsec:nursing-anomaly}

An unexpected finding emerges from our analysis.
The Bachelor of Nursing (BNur) graduates, when combined with AI-900 certification, demonstrate higher skill similarity to data and AI roles than BEng or BCS (ESD) graduates. This counterintuitive result challenges conventional assumptions about technology career pathways and warrants more detailed investigation.

Several interconnected factors contribute to this phenomenon. The most significant relates to skill set size differences across degree programmes. The BNur comprises only $68$ unique skills, substantially fewer than BEng ($491$ skills), BCS (ESD) ($256$ skills), and BCS (AI) ($135$ skills). This size disparity creates a proportional impact whereby AI certification skills represent a much larger proportion of nursing graduates' total skill set after augmentation, as described by our skill combination methodology in \cref{subsec:combine-skillsets}.

Our mathematical framework (\cref{subsec:quantify-certification-impact}) explains this effect.
When certification skills align well with job requirements, smaller baseline skill sets experience more dramatic relative improvements. For nursing graduates, the AI-900 certification effectively doubles their technical skill repertoire, while for engineering or computer science graduates, the same certification represents a smaller proportional enhancement to an already substantial skill base.

Unexpectedly, certain nursing skills demonstrate meaningful alignment with data and AI positions. "FourGen Computer-Aided Software Engineering (CASE) Tools" appears in $31$ times in BNur subjects within our dataset, reflecting the increasing digitalisation of healthcare systems. Similarly, ``Critical Thinking'' appears 14 times, representing a transferable analytical skill highly valued in data science roles. These findings suggest that modern nursing education incorporates more technological and analytical components than traditional perceptions might indicate.

To understand the specific contributions of this skill combination, we examine the latent skill alignment between BNur plus AI-900 and various AI roles. \cref{tab:nursing-skills} presents the top five skills by alignment score for each role when considering the combined nursing and AI-900 skill set, using the alignment approach defined in~\cref{subsec:skill-alignement}.

The analysis reveals three important insights. First, the highest alignment scores consistently derive from technical skills introduced through the AI-900 certification. ``Azure Machine Learning'' achieves an alignment score of $222.91$ for Machine Learning Engineer roles, demonstrating how industry certifications can rapidly introduce highly relevant technical competencies to non-technical degree backgrounds.

Second, healthcare-specific knowledge demonstrates unexpected transferability to technology roles. 
``Cardiac Monitoring'' appears as a relevant skill for Data Analyst positions with a small, but meaninful alignment score of $1.55$, illustrating how domain expertise from nursing can transfer to data-focused positions. This suggests that healthcare professionals possess valuable analytical and monitoring skills that translate effectively to data science contexts, particularly in health informatics and medical data analysis.

Third, nursing education's emphasis on quantitative analysis and precise measurement provides a solid foundation for data-oriented roles. ``Applied Mathematics'' achieves an alignment score of $7.53$ for Data Analyst positions, reflecting the statistical and mathematical components inherent in modern nursing education. This quantitative foundation, combined with critical thinking skills developed through clinical practice, creates an unexpectedly strong basis for data science competencies.

This finding challenges fundamental assumptions about which degrees benefit most from industry certifications and suggests promising alternative pathways into AI careers. Healthcare professionals who combine domain expertise with targeted technical certifications may represent an untapped talent pool for technology roles, particularly in sectors where healthcare knowledge provides competitive advantage. The results suggest that universities and industry partners should consider developing specialised technology tracks within healthcare programmes, while employers should recognise the potential in candidates from non-traditional backgrounds who have obtained relevant certifications.

\section{Discussion, Limitations and Future Work}

Our analysis reveals fundamental insights about how industry certifications can transform educational pathways and employment outcomes. 
The findings challenge traditional assumptions about technology career preparation and suggest concrete strategies for addressing Australia's critical skills shortage.

\subsection{Transforming Educational Programs}

The quantified impact of industry certifications provides compelling evidence for educational reform. Our results demonstrate that strategic integration of industry-applied skills can dramatically enhance graduate employability, particularly when implemented thoughtfully across diverse disciplines.

\textbf{Curriculum Integration Strategies.} 
Universities could move beyond viewing certifications as supplementary credentials and instead embed them as integral components of degree programs. The dramatic improvements observed across all degree types—ranging from 9\% to over 9000\% increases in job alignment—indicate that certification integration represents a systematic opportunity rather than a marginal enhancement. This integration should be particularly prioritised in non-technical disciplines, where our analysis shows the most substantial relative gains.

\textbf{Cross-Disciplinary Innovation.} The nursing anomaly reveals untapped potential for technology career pathways from unexpected educational backgrounds. The finding that nursing graduates combined with AI-900 certification outperformed engineering graduates in several AI roles suggests universities should develop specialised technology tracks within traditionally non-technical programs. Healthcare informatics, educational technology, and business analytics represent natural interdisciplinary bridges where domain expertise combined with targeted technical certifications can create unique value propositions.

\textbf{Data-Driven Curriculum Development.} Our skill decomposition methodology provides educational institutions with precise tools for identifying and addressing skill gaps. Rather than relying on industry feedback or graduate surveys, institutions can now quantify specific competency deficiencies and design targeted interventions. This approach enables continuous curriculum optimisation based on evolving job market requirements, ensuring graduates remain competitive in rapidly changing technological landscapes.

\subsection{Career Development}

Our findings reshape how individuals should approach career development and transition planning in technology sectors.

\textbf{Evidence-Based Certification Selection.} The traditional approach of pursuing prestigious or popular certifications should be replaced with strategic, data-driven selection based on quantified skill alignment with target roles. Our analysis demonstrates that certification value varies dramatically depending on educational background and career objectives. For instance, while AI-900 provides modest improvements for computer science graduates ($23\%$ for Machine Learning Engineer roles), it delivers extraordinary enhancements for nursing graduates ($9296\%$ for the same roles).

\textbf{Alternative Career Pathways.} The research validates non-traditional routes into technology careers, particularly for healthcare and other domain specialists seeking technology transitions. The discovery that nursing education provides unexpected foundations for data analysis roles—through critical thinking, applied mathematics, and systematic measurement approaches—suggests that career changers should leverage rather than abandon their domain expertise when entering technology fields.

\textbf{Targeted Skill Investment.} Our granular skill alignment analysis enables individuals to identify specific competency gaps and invest learning resources strategically. Rather than pursuing broad technology education, professionals can focus on targeted certifications that address precise weaknesses in their skill profiles relative to desired roles.

\subsection{Industry Transformation and Talent Acquisition}

The implications for employers and industry stakeholders extend beyond traditional hiring practices to fundamental rethinking of talent identification and development.

\textbf{Expanding Talent Pipelines.} Our findings provide quantitative evidence for recognising potential in candidates from non-traditional educational backgrounds who have obtained relevant certifications. This is particularly critical given Australia's projected need for $52,000$ new technology professionals annually by 2030. By broadening talent recognition beyond computer science and engineering graduates, organisations can access previously overlooked candidate pools while addressing diversity and inclusion objectives.

\textbf{Skills-Based Assessment.} The research supports shifting hiring practices from credential-based screening to skill alignment assessment. Our methodology enables organisations to evaluate candidates based on quantified competency matches rather than degree prestige or traditional educational pathways. This approach can improve hiring effectiveness while reducing bias against non-traditional candidates.

\textbf{Precision Training Programs.} Organisations can leverage our skill decomposition analysis to design targeted upskilling programs that address specific competency weaknesses. Rather than implementing broad training initiatives, companies can develop precise interventions that maximize return on training investment while accelerating employee transition into high-demand roles.

\subsection{Broader Implications for Skills Policy}

The research contributes to broader discussions about education policy and workforce development in Australia's evolving economy.

\textbf{Public-Private Collaboration.} The demonstrated value of industry certifications suggests enhanced collaboration opportunities between educational institutions and technology companies. Government policy could incentivise such partnerships through funding mechanisms that support certification integration in degree programs, particularly in universities serving diverse student populations.

\textbf{Workforce Transition Support.} The finding that career changers from healthcare and other disciplines can successfully transition into technology roles with targeted certification support suggests opportunities for government-funded reskilling programs. These programs could specifically target professionals in declining industries or those seeking career advancement, providing pathways into high-demand technology roles.

\textbf{Regional Development.} The accessibility of industry certifications, particularly foundational programs like AI-900 that require minimal prerequisites, suggests opportunities for technology workforce development in regional areas where traditional computer science education may be limited. This could contribute to addressing both regional economic development and national skills shortage challenges.

\subsection{Addressing Limitations and Future Considerations}

While our findings provide robust evidence for certification value, several methodological limitations merit consideration for practical implementation and future research directions.

\textbf{Methodological Constraints.} Our reliance on existing skill taxonomies may not capture rapidly emerging competencies, and the static nature of our model does not account for evolving skill relationships over time. Additionally, skill similarity alone represents one dimension of job readiness—factors such as communication abilities, cultural fit, and leadership potential remain important considerations that complement quantitative skill alignment measures.

\textbf{Geographic and Temporal Scope.} This study focuses mainly on the Australian market.
This suggests that findings may require validation in other geographic contexts before broad generalisation, though our methodology provides a transferable framework for similar analyses in different labour markets. The mathematical framework also reveals that irrelevant certifications can theoretically decrease job alignment, emphasizing the critical importance of strategic selection rather than credential accumulation.

\textbf{Future Research Priorities.} Key extensions include multivariate optimisation incorporating variables beyond skill similarity, longitudinal analyses tracking employment outcomes to validate predictive power, and broader certification analysis encompassing AWS, Google, and Cisco credentials. We are particularly focused on developing personalised recommendation systems that leverage predictive analytics and real-time market analysis to guide strategic certification selection based on individual academic backgrounds and career aspirations.

\bibliographystyle{plainnat}

\appendix

\section{Mathematical Derivations}
\label{app:math-derivations}

This appendix provides detailed mathematical derivations that underpin the methodology presented in Section 3. These derivations explain the formal relationship between skill set similarity before and after augmentation with industry certifications.

\subsection{Derivation of the Skill Set Similarity Transformation}

Here we derive the relationship between the similarity of a degree skill set $S_1$ to a job role skill set $T$, and the similarity of the combined degree-certification skill set $S_1'$ to the same job role.

We start with the following expressions for the similarity between skill sets:

\begin{equation}
    \Theta(S_1, T) = \frac{1}{C} \sum_{s_1 \in S_1} \sum_{t \in T} w(s_1, S_1) \cdot w(t, T) \cdot \theta(s_1, t)
\end{equation}

Where $C$ is the normalisation constant:
\begin{equation}
    C = \sum_{s_1 \in S_1} \sum_{t \in T} w(s_1, S_1) \cdot w(t, T)
\end{equation}

Similarly, for the augmented skill set $S_1'$:
\begin{equation}
    \Theta(S_1', T) = \frac{1}{C'} \sum_{s' \in S_1'} \sum_{t \in T} w(s', S_1') \cdot w(t, T) \cdot \theta(s', t)
\end{equation}

With normalisation constant:
\begin{equation}
    C' = \sum_{s' \in S_1'} \sum_{t \in T} w(s', S_1') \cdot w(t, T)
\end{equation}

To derive the relationship between $\Theta(S_1', T)$ and $\Theta(S_1, T)$, we partition the skills in $S_1'$ based on their origin and expand the summation:

\begin{align}
\sum_{s' \in S_1'} \sum_{t \in T} w(s', S_1') \cdot w(t, T) \cdot \theta(s', t) = \nonumber \\
\sum_{s \in S_1, s \notin S_2} \sum_{t \in T} w(s, S_1) \cdot w(t, T) \cdot \theta(s, t) + \nonumber \\
\sum_{s \in S_2, s \notin S_1} \sum_{t \in T} w^T(s, S_2) \cdot w(t, T) \cdot \theta(s, t) + \nonumber \\
\sum_{s \in S_1 \cap S_2} \sum_{t \in T} [w(s, S_1) + w^T(s, S_2)] \cdot w(t, T) \cdot \theta(s, t)
\end{align}

The last term can be expanded as:
\begin{align}
\sum_{s \in S_1 \cap S_2} \sum_{t \in T} [w(s, S_1) + w^T(s, S_2)] \cdot w(t, T) \cdot \theta(s, t) = \nonumber \\
\sum_{s \in S_1 \cap S_2} \sum_{t \in T} w(s, S_1) \cdot w(t, T) \cdot \theta(s, t) + \nonumber \\
\sum_{s \in S_1 \cap S_2} \sum_{t \in T} w^T(s, S_2) \cdot w(t, T) \cdot \theta(s, t)
\end{align}

Substituting this back and rearranging, we get:
\begin{align}
\sum_{s' \in S_1'} \sum_{t \in T} w(s', S_1') \cdot w(t, T) \cdot \theta(s', t) = \nonumber \\
\sum_{s \in S_1} \sum_{t \in T} w(s, S_1) \cdot w(t, T) \cdot \theta(s, t) + \nonumber \\
\sum_{s \in S_2} \sum_{t \in T} w^T(s, S_2) \cdot w(t, T) \cdot \theta(s, t)
\end{align}

Recognising that the first term is equal to $C \cdot \Theta(S_1, T)$, we get:
\begin{equation}
\sum_{s' \in S_1'} \sum_{t \in T} w(s', S_1') \cdot w(t, T) \cdot \theta(s', t) = C \cdot \Theta(S_1, T) + \sum_{s \in S_2} \sum_{t \in T} w^T(s, S_2) \cdot w(t, T) \cdot \theta(s, t)
\end{equation}

Dividing both sides by $C'$, we obtain:
\begin{equation}
\Theta(S_1', T) = \frac{C}{C'} \cdot \Theta(S_1, T) + \frac{1}{C'} \sum_{s \in S_2} \sum_{t \in T} w^T(s, S_2) \cdot w(t, T) \cdot \theta(s, t)
\end{equation}

Similarly, for the normalisation constants, we have:
\begin{align}
C' &= \sum_{s' \in S_1'} \sum_{t \in T} w(s', S_1') \cdot w(t, T) \nonumber \\
&= \sum_{s \in S_1} \sum_{t \in T} w(s, S_1) \cdot w(t, T) + \sum_{s \in S_2} \sum_{t \in T} w^T(s, S_2) \cdot w(t, T) \nonumber \\
&= C + \sum_{s \in S_2} \sum_{t \in T} w^T(s, S_2) \cdot w(t, T)
\end{align}

Since the second term is non-negative (weights are always non-negative), we have $C' \geq C$, which means $\frac{C}{C'} \leq 1$.

\subsection{Special Cases Analysis}

The derived relationship allows us to analyse two important special cases:

\begin{enumerate}
\item \textbf{Case 1: Highly Relevant Certification}

When the certification skills are highly relevant to the target job role, the similarity $\theta(s, t)$ is high for many pairs $(s,t)$ where $s \in S_2$ and $t \in T$. In this case, the second term becomes significant:

\begin{equation}
\frac{1}{C'} \sum_{s \in S_2} \sum_{t \in T} w^T(s, S_2) \cdot w(t, T) \cdot \theta(s, t) \gg 0
\end{equation}

Even though the first term $\frac{C}{C'} \cdot \Theta(S_1, T)$ is scaled down slightly (since $\frac{C}{C'} < 1$), the large positive contribution from the second term ensures that:

\begin{equation}
\Theta(S_1', T) > \Theta(S_1, T)
\end{equation}

This explains why relevant certifications significantly improve skill alignment with target jobs.

\item \textbf{Case 2: Irrelevant Certification}

When certification skills are completely unrelated to the target job role, the similarity $\theta(s, t) \approx 0$ for most pairs $(s,t)$ where $s \in S_2$ and $t \in T$. In this case, the second term approaches zero:

\begin{equation}
\frac{1}{C'} \sum_{s \in S_2} \sum_{t \in T} w^T(s, S_2) \cdot w(t, T) \cdot \theta(s, t) \approx 0
\end{equation}

However, the normalisation constant $C'$ still increases due to the addition of new skills, causing the first term to be scaled down:

\begin{equation}
\Theta(S_1', T) \approx \frac{C}{C'} \Theta(S_1, T) < \Theta(S_1, T)
\end{equation}

This mathematical result explains the counter-intuitive finding that irrelevant certifications can actually decrease one's alignment with target job roles.
\end{enumerate}

\subsection{Implications for Different Educational Backgrounds}

The derivation also explains why non-technical degrees experience larger relative improvements than technical degrees when augmented with the same certification:

\begin{enumerate}
\item For a non-technical degree like nursing:
   \begin{enumerate}
   \item The baseline skill overlap with technical roles is minimal, so $\Theta(S_1, T)$ is very small
   \item The certification adds highly relevant skills that directly align with the target role
   \item The relative increase is large because the starting point is low
   \end{enumerate}

\item For a technical degree like Computer Science:
   \begin{enumerate}
   \item The baseline skill overlap is already substantial, so $\Theta(S_1, T)$ is relatively high
   \item Many certification skills may already be present in the degree curriculum
   \item The relative increase is smaller because the starting point is higher
   \end{enumerate}
\end{enumerate}

This mathematical framework provides the theoretical foundation for our empirical findings and offers guidance for certification selection based on educational background.

\section{Vectorised Skills Space Method (VSSM)}
\label{app-subsec:vssm}

The Vectorised Skills Space Method (VSSM) is an optimised implementation of the Skills Space Method that leverages matrix operations to achieve significant performance improvements while maintaining mathematical equivalence to the original formulation. This section details the implementation approach developed to handle the computational demands of large-scale skill similarity analysis.

\subsection{Motivation for Vectorisation}

A naive implementation of the Skills Space Method involves constructing a set of jobs $J$, where each job $j \in J$ contains a set of skills $S_j$. Computing Skill Set Similarity using nested loops and iterative operations results in poor runtime performance that does not scale with dataset size.

Performance benchmarks of the naive implementation demonstrate polynomial growth in execution time as the number of occupations and skills increases linearly:

\begin{table}[h]
\centering
\begin{tabular}{ccc}
\toprule
\textbf{Unique Occupations} & \textbf{Unique Skills} & \textbf{Mean Execution Time (ms)} \\
\midrule
15 & 5 & 13.47 \\
30 & 10 & 168.04 \\
45 & 15 & 2,805.92 \\
60 & 20 & 10,544.54 \\
75 & 25 & 25,044.62 \\
90 & 30 & 68,196.62 \\
\bottomrule
\end{tabular}
\caption{Performance degradation of naive Skills Space Method implementation}
\end{table}

This exponential growth in computation time necessitates an optimised approach suitable for large datasets containing millions of job advertisements and thousands of unique skills.

\subsection{Data Representation: Binary Presence Matrix}

The foundation of VSSM is transforming the job-skill relationship from an adjacency list representation to a binary presence matrix. Given a set of jobs $J = \{j_0, j_1, \ldots, j_n\}$ and skills $S = \{s_0, s_1, \ldots, s_m\}$, we construct a skill population matrix $P$:

\begin{equation}
P = \begin{bmatrix}
x_{j_0,s_0} & x_{j_0,s_1} & \cdots & x_{j_0,s_m} \\
x_{j_1,s_0} & x_{j_1,s_1} & \cdots & x_{j_1,s_m} \\
\vdots & \vdots & \ddots & \vdots \\
x_{j_n,s_0} & x_{j_n,s_1} & \cdots & x_{j_n,s_m}
\end{bmatrix}
\end{equation}

where:
\begin{equation}
x_{j,s} = \begin{cases}
1 & \text{if } s \in j \\
0 & \text{if } s \notin j
\end{cases}
\end{equation}

This binary matrix representation enables efficient vectorised operations for subsequent computations.

\subsection{Vectorisation of Revealed Comparative Advantage (RCA)}

The RCA function measures skill importance within job advertisements:

\begin{equation}
\text{RCA}(j,s) = \frac{x(j,s) / \sum_{s' \in S} x(j,s')}{\sum_{j' \in J} x(j',s) / \sum_{j' \in J, s' \in S} x(j',s')}
\end{equation}

To vectorise RCA computation, we precompute auxiliary vectors and scalars:

\textbf{Skills per job vector:}
\begin{equation}
b = \begin{pmatrix}
\sum_{s \in S} x(j_0,s) & \sum_{s \in S} x(j_1,s) & \cdots & \sum_{s \in S} x(j_n,s)
\end{pmatrix}
\end{equation}

\textbf{Jobs per skill vector:}
\begin{equation}
c = \begin{pmatrix}
\sum_{j \in J} x(j,s_0) & \sum_{j \in J} x(j,s_1) & \cdots & \sum_{j \in J} x(j,s_m)
\end{pmatrix}
\end{equation}

\textbf{Total skill occurrences:}
\begin{equation}
d = \sum_{j \in J, s \in S} x(j,s)
\end{equation}

The RCA matrix is then computed as:
\begin{equation}
R = P \cdot (b^{-1})^T \cdot c^{-1} \cdot \frac{1}{d}
\end{equation}

where operations are performed element-wise.

\subsection{Vectorisation of Skill Similarity}

Skill similarity $\theta(s_1, s_2)$ measures the likelihood that two skills appear together in job advertisements:

\begin{equation}
\theta(s_1, s_2) = \frac{\sum_{j \in J} e(j,s_1) \cdot e(j,s_2)}{\max\left(\sum_{j \in J} e(j,s_1), \sum_{j \in J} e(j,s_2)\right)}
\end{equation}

where:
\begin{equation}
e(j,s) = \begin{cases}
1 & \text{if } \text{RCA}(j,s) \geq 1 \\
0 & \text{otherwise}
\end{cases}
\end{equation}

The vectorisation process involves:

\textbf{Step 1: Effective use matrix}
\begin{equation}
E = \text{where}(R \geq 1, 1, 0)
\end{equation}

\textbf{Step 2: Joint frequency matrix}
\begin{equation}
N = E^T E
\end{equation}

\textbf{Step 3: Skill frequency vector}
\begin{equation}
q = \text{diag}(N)
\end{equation}

\textbf{Step 4: Antecedent matrix}
\begin{equation}
A = \max(Q^T, Q)
\end{equation}

where $Q$ is the matrix formed by tiling vector $q$.

\textbf{Step 5: Skill similarity matrix}
\begin{equation}
\Theta = N \cdot A^{-1}
\end{equation}

\subsection{Vectorisation of Skill Set Similarity}

For two skill sets represented as subsets of the job population, skill weights are computed as:

\begin{equation}
w(s,S) = \frac{1}{|J'|} \sum_{j \in J'} \text{RCA}(j,s)
\end{equation}

where $J'$ is the subset of jobs belonging to the skill set.

The vectorised computation proceeds as follows:

\textbf{Step 1: Skill presence vectors}
For each skill set, create a binary vector $S'$ indicating skill presence:
\begin{equation}
S'_i = \begin{cases}
1 & \text{if skill } i \text{ appears in any job in } J' \\
0 & \text{otherwise}
\end{cases}
\end{equation}

\textbf{Step 2: Skill weight computation}
\begin{equation}
w = \frac{1}{|J'|} \sum_{j \in J'} (S' \cdot R_{j,:})
\end{equation}

\textbf{Step 3: Weighted similarity matrix}
\begin{equation}
\Theta_{weighted} = w_2 \cdot w_1^T \cdot \Theta
\end{equation}

\textbf{Step 4: Final skill set similarity}
\begin{equation}
\text{SSS}(S_1, S_2) = \frac{\sum_{i,j} \Theta_{weighted}(i,j)}{\sum_{i,j} W(i,j)}
\end{equation}

where $W = w_1^T \cdot w_2$ is the weight product matrix.

\subsection{Performance Results}

Benchmarking was conducted on two distinct hardware configurations to evaluate VSSM performance across different computational architectures:

\begin{itemize}
\item \textbf{Macbook Pro 2021}: M1 Pro chip (8-core CPU, 10-core GPU), 16GB RAM, using MLX acceleration
\item \textbf{PC Workstation}: AMD Ryzen 5600X (6-core CPU), 32GB RAM, NVIDIA RTX 3060 12GB, using CuPy acceleration
\end{itemize}

Comparative benchmarking demonstrates substantial performance improvements across both platforms:

\begin{table}[h]
\centering
\begin{tabular}{cccccc}
\toprule
\textbf{Test} & \textbf{Occupations} & \textbf{Skills} & \textbf{Baseline (ms)} & \textbf{VSSM (ms)} & \textbf{Speedup} \\
\midrule
1 & 15 & 5 & 11.55 & 0.15 & 77× \\
2 & 30 & 10 & 141.58 & 0.14 & 1,011× \\
3 & 45 & 15 & 2,206.75 & 0.20 & 11,034× \\
4 & 60 & 20 & 5,673.70 & 0.20 & 28,369× \\
5 & 75 & 25 & 28,927.69 & 0.30 & 96,426× \\
6 & 90 & 30 & 34,530.35 & 0.24 & 143,876× \\
\bottomrule
\end{tabular}
\caption{Performance comparison between naive baseline and VSSM implementation on PC workstation}
\end{table}

The performance difference between Apple Silicon and NVIDIA GPU platforms was negligible for smaller datasets (test iterations 1-5), with the NVIDIA GPU system demonstrating approximately 5.5× faster execution for the largest dataset (test iteration 6). Both platforms achieved dramatic improvements over the baseline implementation, with speed enhancements ranging from 77× to 143,876×, making large-scale skill similarity analysis computationally feasible for real-world applications.

\subsection{Implementation Validation}

Mathematical equivalence between the naive and vectorised implementations is verified through regression testing. Both implementations process identical randomly generated datasets, and outputs are compared to ensure numerical accuracy within floating-point precision limits.

\subsection{Computational Infrastructure}

VSSM is implemented using optimised Python libraries:
\begin{itemize}
\item \textbf{NumPy}: For CPU-based matrix operations with BLAS optimisation
\item \textbf{CuPy}: For GPU-accelerated computations on NVIDIA hardware
\item \textbf{MLX}: For Apple Silicon GPU acceleration
\item \textbf{SciPy sparse}: For memory-efficient sparse matrix operations
\end{itemize}

This multi-backend approach ensures optimal performance across different hardware configurations while maintaining a consistent API for skill similarity computations.

\section{Implementation Considerations}
\subsection{Stakeholder-Specific Implementation Guidance}

For effective implementation, different stakeholders should consider:

\paragraph{Universities:}
\begin{itemize}
    \item Integrate certification modules within existing degree programs
    \item Use skill alignment data to inform curriculum updates
    \item Partner with certification providers for streamlined student pathways
\end{itemize}

\paragraph{Employers:}
\begin{itemize}
    \item Implement skill similarity assessments in recruitment processes
    \item Identify internal candidates for role transitions based on skill alignment
    \item Design targeted training programs addressing specific skill weaknesses
\end{itemize}

\paragraph{Certification Providers:}
\begin{itemize}
    \item Align certification design with quantified skill gaps
    \item Develop specialised certification paths for specific degree backgrounds
    \item Provide integration tools for educational institutions
\end{itemize} 
\end{document}